%% file: mode3_regular_arxiv01.tex
\documentclass[10pt]{article}
\usepackage{amssymb,amsbsy}
\usepackage{amsmath}
\usepackage{amsthm}
\usepackage{graphics,graphicx}
\usepackage[T1]{fontenc}
\usepackage{color}

\input{definitions}

\title{Perturbation of Mode III interfacial cracks}

\author{A. Piccolroaz$^{(1)}$, G. Mishuris$^{(2)}$, A.B. Movchan$^{(3)}$
\\
\\
$^{(1)}$
{\it Dipartimento di Ingegneria Meccanica e
Strutturale, Universit\`a di Trento,}
\\ {\it Via Mesiano 77, I-38050 Trento, Italia}
\\
{\it
$^{(2)}$
Institute of Mathematical and Physical Sciences, Aberystwyth University, }
\\ {\it Ceredigion SY23 3BZ, Wales U.K.,}
\\
{\it
$^{(3)}$
Department of Mathematical Sciences, University of Liverpool, }
\\ {\it Liverpool L69 3BX, U.K.}
}

\begin{document}

\maketitle

\begin{abstract}
\noindent
We consider the perturbation problem of a Mode III interfacial crack. The perturbation is of geometrical type and can be both 
perturbation of the crack faces and perturbation of the interface, which can deviate from the initial straight line configuration. 
Asymptotic formulae are derived for the first-order perturbation of the stress intensity factor. It is shown that, due to the 
unsymmetrical nature of the problem, the Mode III skew-symmetric weight function derived in Piccolroaz et al. (2009) is essential for 
the derivation of the correct asymptotic formulae. 

To illustrate the method, we present the numerical results for different geometrical perturbations of a half-plane interfacial 
crack in an infinite bimaterial structure. Discussion on the extension of the method to finite bodies is also presented. 
\end{abstract}

\newpage

\tableofcontents

\newpage

\section{Introduction}
\label{sec1}
Modelling of interfacial cracks was addressed in the fundamental papers by Willis (1971) and Hutchinson et al.\ (1987). 
Oscillatory asymptotics of displacement and stress fields near the crack edges represent an important feature of interfacial 
cracks in vector problems of elasticity. Singular integral equations for the displacement jump across the crack take into 
account the full coupling between the normal and shear modes of loading.

The notion of weight functions for cracks, as stress-intensity factors associated with a point force load, was introduced in 
Bueckner (1985, 1989). Another approach, viewing weight functions as special singular solutions of homogeneous problems for 
cracks, was used in Willis and Movchan (1995). Analysis of weight functions, represented by special singular solutions of 
the homogeneous problem for an interfacial crack was performed in Lazarus and Leblond (1998) and Antipov (1999).
Further studies, involving both symmetric and skew-symmetric weight functions, are included in Bercial-Velez et al.\ (2005) and 
Piccolroaz et al.\ (2007). In particular, as the three-dimensional crack advances, the crack front deviates, and for the case of 
in-plane deviations the paper Piccolroaz et al.\ (2007) incorporates the first-order analysis of the weight functions based 
on the full solution of the corresponding matrix Wiener-Hopf equation. It has been shown by Piccolroaz et al.\ (2009) that the 
lack of symmetry inherent in the interfacial crack problem induces the appearance of a skew-\-symme\-tric component of the weight 
functions, so that, for instance, the skew-symmetric loads generate non-zero stress intensity factors even in the case of two 
dimensions (plane strain, plane stress or antiplane shear).

We note that  the problem of finding the variation of the stress intensity factors induced by a small geometrical perturbation of a
plane crack placed at the interface between two dissimilar elastic materials requires the use of not only the  symmetric weight 
functions but also of the skew-symmetric components. This statement is valid not only for a full vector problem, but it is also true 
for relatively simple anti-plane shear formulations. This is shown in the present paper with reference to the case of Mode III 
deformation.

We consider both perturbation of the crack faces and of the interface, so that, in its perturbed state, the crack faces occupy
the lines
$$
\gamma_\pm^\epsilon = \{(x_1,x_2 ): x_2 = \epsilon \psi_\pm(x_1), -\infty < x_1 < 0\},
$$
whereas the interface occupies the line
$$
\zeta^\epsilon = \{(x_1,x_2 ): x_2 = \epsilon \phi(x_1), 0 < x_1 < \infty\}.
$$
Here, we assume that the  functions $\psi_\pm, \phi$ and their derivatives vanish within a finite neighbourhood of the crack tip. 
Such perturbation is referred to as the {\em regular perturbation of the boundary.}

Using the symmetric and skew-symmetric weight functions for the interfacial crack, we derive asymptotic formulae for the perturbed 
stress intensity factor, which characterize the stress in the vicinity of the crack tip. 

The paper is organized as follows. The formulation of the problem and the governing equations are outlined in Section \ref{sec2}. 
Section \ref{sec3} introduces the symmetric and skew-symmetric weight functions for the interfacial crack; the same section outlines 
the regular perturbation asymptotic procedure. An illustrative example and applications of the asymptotic formulae are included in 
Section \ref{sec4}.

\section{Problem formulation}
\label{sec2}
\begin{figure}[ht]
\begin{center}
\includegraphics[width=7cm]{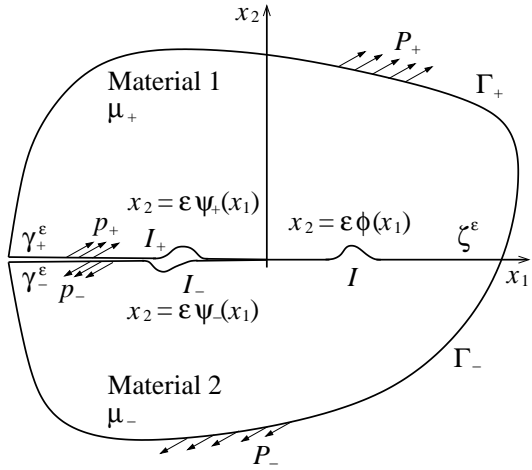}
\caption{\footnotesize Geometry of the problem}
\end{center}
\label{fig01}
\end{figure}

We consider a twodimensional bimaterial structure made of two dissimilar materials, joined along a plane interface, see 
Fig.~\ref{fig01}. The two materials are assumed to be linear elastic and isotropic, with elastic moduli denoted by $\mu_\pm$ 
(shear modulus) and $\nu_\pm$ (Poisson's ratio). A plane crack is placed along the interface, which is assumed to be perfect, so that 
the displacement and traction components are continuous across the interface $\zeta^\epsilon$.

The cracked body is loaded by out-of-plane tractions: $P_\pm = \mu_\pm \partial u^\pm/\partial n^\pm$ along the outer boundary 
$\Gamma_\pm$, and $p_\pm = \mu_\pm \partial u^\pm/\partial x_2$ along the crack faces $\gamma_\pm^\epsilon$, where $u$ denotes the 
displacement component along $x_3$. The loading is self-balanced in terms of both principal force and moment vectors
\beq
\int_{\Gamma_+} P_+ d\bx + \int_{\Gamma_-} P_- d\bx -
\int_{\gamma_+^\epsilon} p_+ d\bx + \int_{\gamma_-^\epsilon} p_- d\bx = 0.
\eeq

Suppose that the plane crack faces and the plane interface are slightly perturbed: the perturbation is denoted by $x_2 = \epsilon \psi_\pm(x_1)$
with supports $I_\pm$ for the crack faces and $x_2 = \epsilon \phi(x_1)$ with support $I$ for the interface, where $\epsilon$ is 
a small parameter. We are interested in the calculation of the corresponding perturbation of the stress intensity factor $K_\modIII$.

%%%%%%%%%%%%%%%%%%%%%%%%%%%%%%%%%%%%%%%%%%%%%%%%%%%%%%%%%%%%%%%%%%%%%%
%%%%%%%%%%%%%%%%%%%%%%%%%%%%%%%%%%%%%%%%%%%%%%%%%%%%%%%%%%%%%%%%%%%%%%

The problem is then formulated in terms of the Laplace equation
\beq
\label{laplace}
\Delta u^\pm(x_1,x_2) = 0,
\eeq
with boundary conditions:
\beq
\label{boundary_external}
\mu_\pm \frac{\partial u^\pm}{\partial n^\pm} = P_\pm \quad \text{on $\Gamma_\pm$},
\eeq
\beq
\label{boundary_crack}
\mu_\pm \frac{\partial u^\pm}{\partial x_2} = p_\pm \quad \text{on $\gamma_\pm^\epsilon$},
\eeq
and interface conditions
\beq
\label{interface}
u^+ = u^-, \quad \mu_+ \frac{\partial u^+}{\partial n} = \mu_- \frac{\partial u^-}{\partial n} \quad \text{on $\zeta^\epsilon$}.
\eeq
We assume that the loading $p_\pm$ along the crack faces $\gamma_\pm^\epsilon$ is separated from the supports $I_\pm$.

\section{Solution of the perturbation problem by means of symmetric and skew-symmetric weight functions}
\label{sec3}
The solution of the perturbed problem, $u^\pm(x_1,x_2,\epsilon)$, can be expanded for small $\epsilon$ as follows
\beq
u^\pm(x_1,x_2,\epsilon) = u_0^{\pm}(x_1,x_2) + \epsilon u_1^{\pm}(x_1,x_2) + O(\epsilon^2),
\eeq
where $u_0^{\pm}(x_1,x_2)$ is the solution of the unperturbed problem ($\epsilon = 0$) and $u_1^{\pm}(x_1,x_2)$ is the first-order 
variation of the solution with respect to $\epsilon$.

\subsection{Model problem for $u_0^{\pm}(x_1,x_2)$}
\label{sec31}
\begin{figure}[ht]
\begin{center}
\includegraphics[width=7cm]{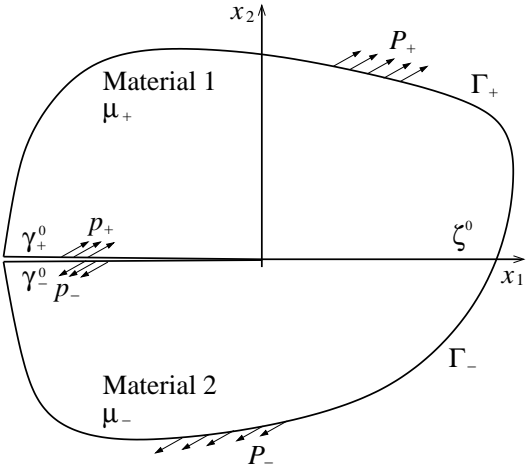}
\caption{\footnotesize Model problem for $u_0^{\pm}(x_1,x_2)$}
\end{center}
\label{fig02}
\end{figure}

The model problem for $u_0^{\pm}(x_1,x_2)$ is obtained by taking the leading order asymptotics of the perturbed problem as $\epsilon=0$, 
so that $u_0^{\pm}(x_1,x_2)$ must satisfy the Laplace equation \eq{laplace} with the boundary and interfacial conditions 
\eq{boundary_external} -- \eq{interface} where the boundaries $\gamma_\pm^\epsilon$ and $\zeta^\epsilon$ are replaced by $\gamma_\pm^0$
and $\zeta^0$, respectively, see Fig. \ref{fig02}.

%%%%%%%%%%%%%%%%%%%%%%%%%%%%%%%%%%%%%%%%%%%%%%%%%%%%%%%%%%%%%%%%%%%%%%
%%%%%%%%%%%%%%%%%%%%%%%%%%%%%%%%%%%%%%%%%%%%%%%%%%%%%%%%%%%%%%%%%%%%%%

The solution with locally bounded elastic energy and vanishing at infinity can be easily found by means of Mellin transform in polar 
coordinates. The Mellin transforms for the displacement vector and the stress tensor  are defined as follows
\beq
\tilde{\bu} = \int_0^\infty \bu(r,\theta) r^{s-1} dr, \quad \tilde{\bsigma} = \int_0^\infty \bsigma(r,\theta) r^s dr,
\eeq
and they are represented by analytic functions in the strips $0< \Re(s) <0.5$ and $-0.5 < \Re(s) < 0.5$, respectively.

Correspondingly, the inverse transforms are 
\beq
\bu(r,\theta) = \frac{1}{2\pi i} \int_{\omega_1 - i\infty}^{\omega_1 +i\infty} \tilde{\bu}(s,\theta) r^{-s} ds, \quad
\bsigma(r,\theta) = \frac{1}{2\pi i} \int_{\omega_2 - i\infty}^{\omega_2 + i\infty} \tilde{\bsigma}(s,\theta) r^{-s - 1} ds,
\eeq
where $\omega_1,\omega_2$ lie in the respective intervals.

With taking into account all boundary conditions, one can obtain the solution in the form (see Appendix \ref{app01} for details):
\beq
\label{u_0} 
\tilde{u}^\pm(s,\theta) = 
\frac{(\tilde{p}_+ - \tilde{p}_-) \cos(s\theta)}{(\mu_+ + \mu_-) s \sin(\pi s)} - 
\frac{(\mu_- \tilde{p}_+ + \mu_+ \tilde{p}_-) \sin(s\theta)}{\mu_\pm (\mu_+ + \mu_-) s \cos(\pi s)}.
\eeq

Introducing the symmetric and skew-symmetric loading, $\jump{0.15}{\tilde{p}} = \tilde{p}_+ - \tilde{p}_-$ and 
$\langle \tilde{p} \rangle = (\tilde{p}_+ + \tilde{p}_-)/2$ respectively, we obtain
\beq
\tilde{u}^\pm(s,\theta) = 
-\frac{\sin(s\theta)}{\mu_\pm s \cos(\pi s)} \langle \tilde{p} \rangle(s) +
\left[ \frac{\cos(s\theta)}{(\mu_+ + \mu_-) s \sin(\pi s)} + 
\frac{(\mu_+ - \mu_-) \sin(s\theta)}{2\mu_\pm (\mu_+ + \mu_-) s \cos(\pi s)} \right] \jump{0.15}{\tilde{p}}(s)
\eeq

The stress components are given by
\beq
\tilde{\sigma}_{3\theta}^\pm(s,\theta) = -
\frac{\cos(s\theta)}{\cos(\pi s)} \langle \tilde{p} \rangle(s) -
\left[ \frac{\mu_\pm \sin(s\theta)}{(\mu_+ + \mu_-) \sin(\pi s)} - 
\frac{(\mu_+ - \mu_-) \cos(s\theta)}{2(\mu_+ + \mu_-) \cos(\pi s)} \right] \jump{0.15}{\tilde{p}}(s),
\eeq
\beq
\tilde{\sigma}_{3r}^\pm(s,\theta) =
\frac{\sin(s\theta)}{\cos(\pi s)} \langle \tilde{p} \rangle(s) - 
\left[ \frac{\mu_\pm \cos(s\theta)}{(\mu_+ + \mu_-) \sin(\pi s)} +
\frac{(\mu_+ - \mu_-) \sin(s\theta)}{2(\mu_+ + \mu_-) \cos(\pi s)} \right] \jump{0.15}{\tilde{p}}(s).
\eeq

This allows to obtain the two-terms asymptotics of tractions ahead of the crack tip and crack opening:
\beq
\sigma_{3\theta}(r,0) = \frac{K_\modIII}{\sqrt{2\pi}} r^{-1/2} + \frac{A_\modIII}{\sqrt{2\pi}} r^{1/2} + O(r^{3/2}),
\eeq
\beq
\jump{0.15}{u}(r) =
\frac{\mu_+ + \mu_-}{\mu_+\mu_-} \left(\frac{2K_\modIII}{\sqrt{2\pi}} r^{1/2} - \frac{2A_\modIII}{3\sqrt{2\pi}} r^{3/2}\right) + O(r^{5/2}),
\eeq
respectively, where
\beq
\label{K_30}
K_\modIII = -\sqrt{\frac{2}{\pi}} \int_0^\infty \left\{ \langle p \rangle(r) + \frac{\eta}{2} \jump{0.15}{p}(r) \right\} r^{-1/2} dr,
\eeq
\beq
A_\modIII = \sqrt{\frac{2}{\pi}} \int_0^\infty \left\{ \langle p \rangle(r) + \frac{\eta}{2} \jump{0.15}{p}(r) \right\} r^{-3/2} dr,
\eeq
in which $\eta = (\mu_- - \mu_+)/(\mu_+ + \mu_-)$ is the contrast parameter.

\subsection{Model problem for $u_1^{\pm}(x_1,x_2)$}
\label{sec32}
To obtain the model problem for $u_1^{\pm}(x_1,x_2)$ we will follow the procedure described in Movchan and Movchan (1995) and 
expand the term $\frac{\partial u^\pm}{\partial n^\pm}$ along $\gamma_\pm^\epsilon$ and $\zeta^\epsilon$, and also $u^{\pm}$ 
along $\zeta^\epsilon$.

We obtain:
\beq
\left. \frac{\partial u^\pm}{\partial n^\pm} \right|_{\gamma_\pm^\epsilon} = 
\left. \frac{\partial u_0^{\pm}}{\partial x_2} \right|_{x_2=0^\pm} + 
\epsilon \left\{ \left. \frac{\partial u_1^{\pm}}{\partial x_2} \right|_{x_2=0^\pm} - 
\frac{\partial}{\partial x_1} \left( \psi_\pm(x_1) \left.
\frac{\partial u_0^{\pm}}{\partial x_1} \right|_{x_2=0^\pm} \right)
\right\} + O(\epsilon^2), \quad x_1 < 0,
\eeq
\beq
\left. \frac{\partial u^\pm}{\partial n} \right|_{\zeta^\epsilon} = 
\left. \frac{\partial u_0^{\pm}}{\partial x_2} \right|_{x_2=0^\pm} +
\epsilon \left\{ \left. \frac{\partial u_1^{\pm}}{\partial x_2} \right|_{x_2=0^\pm} - 
\frac{\partial}{\partial x_1} \left( \phi(x_1) \left.
\frac{\partial u_0^{\pm}}{\partial x_1} \right|_{x_2=0^\pm} \right)
\right\} + O(\epsilon^2), \quad x_1 > 0,
\eeq
and
\beq
u^\pm \Big|_{\zeta^\epsilon} = 
u_0^{\pm}(x_1,0) + 
\epsilon \left\{ 
u_1^{\pm}(x_1,0) + \phi(x_1) \left. \frac{\partial u_0^{\pm}}{\partial x_2} \right|_{x_2=0^\pm} 
\right\} + O(\epsilon^2), \quad x_1 > 0.
\eeq

%%%%%%%%%%%%%%%%%%%%%%%%%%%%%%%%%%%%%%%%%%%%%%%%%%%%%%%%%%%%%%%%%%%%%%
\begin{figure}[ht]
\begin{center}
\includegraphics[width=7cm]{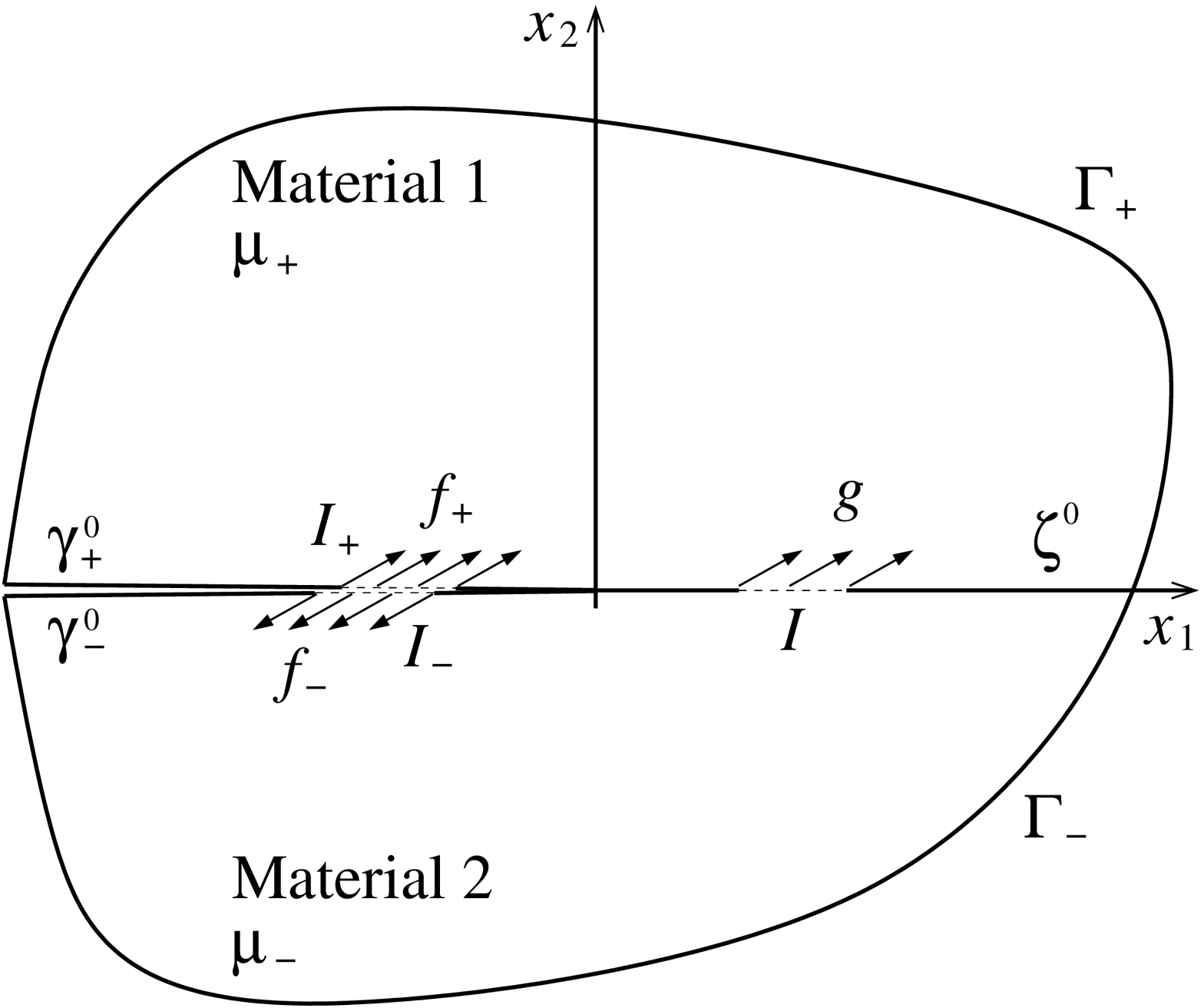}
\caption{\footnotesize Model problem for $u_1^{\pm}(x_1,x_2)$}
\end{center}
\label{fig03}
\end{figure}
%%%%%%%%%%%%%%%%%%%%%%%%%%%%%%%%%%%%%%%%%%%%%%%%%%%%%%%%%%%%%%%%%%%%%%

Therefore the model problem for $u_1^{\pm}(x_1,x_2)$ (see Fig. \ref{fig03}) is formulated in terms of the Laplace equation 
\eq{laplace} with boundary conditions
\beq
\label{boundary_external_new}
\mu_\pm \frac{\partial u_1^{\pm}}{\partial n^\pm} = 0 ,
\eeq
along the boundary $\Gamma_\pm$,
\beq
\label{boundary_crack_new}
\mu_\pm \frac{\partial u_1^{\pm}}{\partial x_2} = \frac{\partial}{\partial x_1} f_\pm(x_1) := 
\mu_\pm \frac{\partial}{\partial x_1} \left( \psi_\pm(x_1) \frac{\partial u_0^{\pm}}{\partial x_1} \right),
\eeq
along $\gamma_\pm^0$, and the interfacial conditions
\beq
\label{boundary_interface_1}
u_1^{+} - u_1^{-} = g(x_1) := -\phi(x_1) \left( \frac{\partial u_0^+}{\partial x_2} - \frac{\partial u_0^-}{\partial x_2} \right) =
-\phi(x_1) \frac{\partial u_0^+}{\partial x_2} \left( 1 - \frac{\mu_+}{\mu_-} \right), \quad \text{on $\zeta^0$},
\eeq
% \vspace*{3mm}
\beq
\label{boundary_interface_2}
\mu_+ \frac{\partial u_1^{+}}{\partial x_2} - \mu_- \frac{\partial u_1^{-}}{\partial x_2} = h(x_1) := 
\mu_+ \frac{\partial}{\partial x_1} \left( \phi(x_1) \frac{\partial u_0^{+}}{\partial x_1} \right) - 
\mu_- \frac{\partial}{\partial x_1} \left( \phi(x_1) \frac{\partial u_0^{-}}{\partial x_1} \right), 
\quad \text{on $\zeta^0$}.
\eeq

Note that the right-hand sides in \eq{boundary_crack_new}, \eq{boundary_interface_2} can be interpreted as ``effective'' loading 
along the crack faces and along the interface, respectively, whereas \eq{boundary_interface_1} is a prescribed discontinuity of 
displacement along the interface.

\subsection{The reciprocity identity and evaluation of the stress intensity factor}
\label{sec33}
The integral representation of the $K_\modIII$ stress intensity factor can be obtained by application of the Betti identity, in 
a way similar to Piccolroaz et al. (2007). The formula is extended here to include the case of imperfect interface with prescribed 
discontinuities of displacement $u$ and traction $\sigma = \mu \partial u/\partial x_2$.

The reciprocity identity reads
\beq
\label{recid}
\int_{-\infty}^\infty \Big\{ \jump{0.15}{U}(x_1'-x_1) \langle \sigma \rangle(x_1) + \langle U \rangle(x_1'-x_1) \jump{0.15}{\sigma}(x_1)
- \langle \Sigma \rangle(x_1'-x_1) \jump{0.15}{u}(x_1) \Big\} dx_1 = 0,
\eeq
where $\jump{0.15}{U}$ and $\langle U \rangle$ are the symmetric and skew-symmetric Mode III weight functions and $\langle \Sigma \rangle$ 
the corresponding traction (see Piccolroaz et al., 2009),
\beq
\label{wfIII}
\jump{0.15}{U}(x_1) = 
\left\{ 
\barr{ll}
\ds \frac{1-i}{\sqrt{2\pi}} x_1^{-1/2}, & \ds x_1 > 0, \\[3mm]
0, & x_1 < 0, 
\earr 
\right.
\eeq
\beq
\label{wfIII2}
\langle \Sigma \rangle(x_1) = 
\left\{
\barr{ll}
0, & \ds x_1 > 0, \\[3mm]
\ds \frac{(1-i) \mu_+\mu_-}{2\sqrt{2\pi}(\mu_+ + \mu_-)} (-x_1)^{-3/2}, & x_1 < 0,
\earr
\right.
\eeq
and $\langle U \rangle(x_1) = \eta/2 \jump{0.15}{U}(x_1)$.

Let us introduce the notations
$$ 
f^{(+)}(x_1) = f(x_1) H(x_1), \quad f^{(-)}(x_1) = f(x_1) H(-x_1), 
$$
where $H$ denotes the Heaviside function, so that
\beq
f(x_1) = f^{(+)}(x_1) + f^{(-)}(x_1).
\eeq

The reciprocity identity \eq{recid} can be written as
$$
\int_{-\infty}^{\infty} \left\{
\jump{0.15}{U}(x_1'-x_1) \langle \sigma \rangle^{(+)}(x_1)
- \langle \Sigma \rangle(x_1'-x_1) \jump{0.15}{u}^{(-)}(x_1)
\right\} dx_1 =
$$
\beq
\label{recid2}
- \int_{-\infty}^{\infty} \left\{
\jump{0.15}{U}(x_1'-x_1) \langle \sigma \rangle^{(-)}(x_1)
+ \langle U \rangle(x_1'-x_1) \jump{0.15}{\sigma}^{(-)}(x_1)
\right\} dx_1
\eeq
$$
- \int_{-\infty}^{\infty} \langle U \rangle(x_1'-x_1)
\jump{0.15}{\sigma}^{(+)}(x_1) dx_1
+ \int_{-\infty}^{\infty} \langle \Sigma \rangle(x_1'-x_1)
\jump{0.15}{u}^{(+)}(x_1) dx_1.
$$

Note that $\langle \sigma \rangle^{(-)}$, $\jump{0.15}{\sigma}^{(-)}$ are average and jump of the prescribed loading on the crack 
faces, whereas $\jump{0.15}{\sigma}^{(+)}$, $\jump{0.15}{u}^{(+)}$ are the prescribed discontinuities of traction and displacement 
along the interface.

The asymptotic procedure described in Piccolroaz et al. (2009) allows us to obtain the integral formula for the computation of 
$K_\modIII$ as
$$
K_\modIII =
-(1+i) \lim_{x_1' \to 0} \int_{-\infty}^{0} \left\{ \jump{0.15}{U}(x_1'-x_1)\langle \sigma \rangle^{(-)}(x_1) + 
\langle U \rangle(x_1'-x_1) \jump{0.15}{\sigma}^{(-)}(x_1) \right\} dx_1 -
$$
\beq
(1+i) \lim_{x_1' \to 0} \int_{0}^{\infty} \langle U \rangle(x_1' - x_1) \jump{0.15}{\sigma}^{(+)}(x_1) dx_1 
+(1+i) \lim_{x_1' \to 0} \int_{0}^{\infty} \langle \Sigma \rangle(x_1' - x_1) \jump{0.15}{u}^{(+)}(x_1) dx_1,
\eeq
which, upon substitution of \eq{wfIII} and \eq{wfIII2}, simplifies to
$$
K_\modIII = \frac{\mu_+\mu_-}{\sqrt{2\pi}(\mu_+ + \mu_-)} \int_{0}^{\infty} \jump{0.15}{u}^{(+)}(x_1) x_1^{-3/2} dx_1 - 
$$
\beq
\label{SIF}
\sqrt{\frac{2}{\pi}} \int_{-\infty}^{0} \left\{ \langle \sigma \rangle^{(-)}(x_1) + 
\frac{\eta}{2} \jump{0.15}{\sigma}^{(-)}(x_1) \right\} (-x_1)^{-1/2} dx_1.
\eeq

Note from \eq{SIF} that the traction discontinuity $\jump{0.15}{\sigma}^{(+)}$, prescribed along the interface, does not contribute 
to the stress intensity factor. This is because the weight function $\langle U \rangle$ is identically zero for $x_1 < 0$.

Now, we consider the asymptotic expansion of $K_\modIII$ with respect to $\epsilon$ as follows
\beq
K_\modIII = K_\modIII^0 + \epsilon K_\modIII^1 + O(\epsilon^2),
\eeq
and make use of the integral formula \eq{SIF} to compute both the leading term $K_\modIII^0$ and the first order variation $K_\modIII^1$. 
It is evident that the leading term $K_\modIII^0$ corresponds to the unperturbed problem ($\epsilon = 0$) and was defined in (\ref{K_30}).

The second term $K_\modIII^1$ corresponds to the model problem described in Sec. \ref{sec32} and can be written as
\beq
K_\modIII^1 = K_\modIII^{1(a)} + K_\modIII^{1(b)},
\eeq
where $K_\modIII^{1(a)}$ and $K_\modIII^{1(b)}$ denote the variations due to the perturbation of the crack faces and of the interface, 
respectively.

Using \eq{SIF} with $\sigma^{(-)} = f$, and $\jump{0.15}{u}^{(+)} = g$, we obtain
\beq
\label{SIF_1a} 
K_\modIII^{1(a)} = 
-\sqrt{\frac{2}{\pi}} \int_{-\infty}^0 
(-x_1)^{-1/2}\frac{\partial}{\partial x_1}\left\{ \langle f \rangle(x_1) + \frac{\eta}{2} \jump{0.15}{f}(x_1) \right\} dx_1,
\eeq
and
\beq
\label{SIF_1b}
K_\modIII^{1(b)} = \frac{\mu_+\mu_-}{\sqrt{2\pi}(\mu_+ + \mu_-)} \int_{0}^{\infty} g(x_1) x_1^{-3/2} dx_1
\eeq
where both jump and average notations are standard and
\beq
\langle f \rangle = \left\langle \mu \psi(x_1) \frac{\partial u_0}{\partial x_1} \right\rangle, \quad
\jump{0.15}{f} = \bjump{0.55}{\mu \psi(x_1) \frac{\partial u_0}{\partial x_1}}, \quad 
g(x_1) = -\phi(x_1) \bjump{0.55}{\frac{\partial u_0}{\partial x_2}},
\eeq
in which the supports of the functions belong to the corresponding boundaries $\gamma_\pm^0$ and $\zeta^0$.

Integrating by parts the first integral \eq{SIF_1a} we get
$$
K_\modIII^{1(a)} = \frac{1}{\sqrt{2\pi}} \int_{-\infty}^{0}
\left\{
\left\langle \mu u_0(x_1) \frac{\partial}{\partial x_1} \left( (-x_1)^{-3/2} \psi(x_1) \right) \right\rangle + \right.
$$
\beq
\label{k1a}
\left. \frac{\eta}{2} \bjump{0.55}{\mu u_0(x_1) \frac{\partial}{\partial x_1} \left( (-x_1)^{-3/2} \psi(x_1) \right)} \right\} dx_1,
\eeq
in which, from the solution described in (\ref{u_0}), we can write for $x_1 < 0$:
\beq
\label{int1a}
\mu_\pm u_0^\pm(x_1) = 
\frac{1}{2\pi i} \int_{-i\infty+\delta}^{i\infty+\delta} 
\left\{ \left[ \frac{1 \mp \eta}{2} \frac{\cot \pi s}{s} \mp \frac{\eta}{2} \frac{\tan \pi s}{s} \right] \jump{0.15}{\tilde{p}}(s) \mp 
\frac{\tan \pi s}{s} \langle \tilde{p} \rangle(s) \right\} (-x_1)^{-s} ds,
\eeq
where $0 < \delta < 1/2$, and for $x_1 > 0$:
\beq
\label{int1b}
\frac{\partial u_0^\pm}{\partial x_2}(x_1) =
-\frac{1}{2\pi i \mu_\pm} \int_{-i\infty}^{i\infty} 
\left\{ \frac{\eta}{2} \jump{0.15}{\tilde{p}}(s) + \langle \tilde{p} \rangle(s) \right\} \frac{x_1^{-s-1}}{\cos \pi s} ds.
\eeq

\section{An illustrative example}
\label{sec4}
\begin{figure}[h]
\begin{center}
\includegraphics[width=7cm]{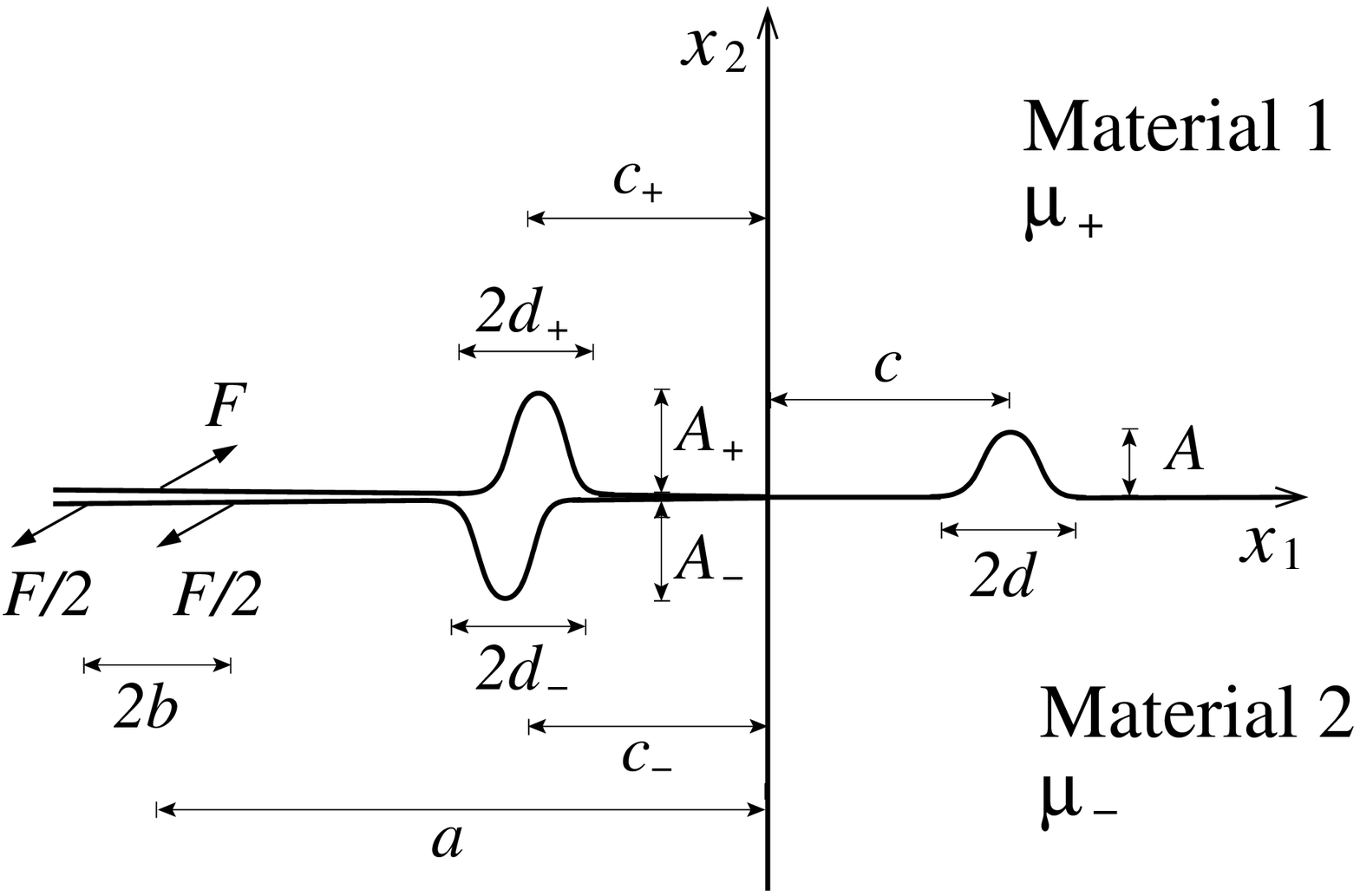}
\caption{\footnotesize Illustrative example}
\end{center}
\label{fig04}
\end{figure}

To illustrate the perturbation method described above, we show in this section numerical results concerning the perturbation of both 
crack faces and interface geometry in the case of an interfacial half-plane crack in an infinite bimaterial structure, see Fig. \ref{fig04}. 
The crack faces are loaded by a ``three-point'' loading system consisting of: a point force $F$ acting upon the upper crack face at a 
distance $a$ behind the crack tip and two point forces $F/2$ acting upon the lower crack face at a distance $a - b$ and $a + b$ behind 
the crack tip, see Fig. \ref{fig04}. In terms of the Dirac delta function $\delta(\cdot)$, the loading is then given by
\beq
p_+(x_1) = F \delta(x_1 + a),
\eeq
\beq
p_-(x_1) = \frac{F}{2} \delta(x_1 + a + b) + \frac{F}{2} \delta(x_1 + a - b),
\eeq
or, in terms of symmetric and skew-symmetric parts, by
$$
\langle p \rangle(x_1) = \frac{1}{2} \delta(x_1 + a) + \frac{1}{4} \delta(x_1 + a + b) + \frac{1}{4} \delta(x_1 + a - b),
$$
$$
\jump{0.15}{p}(x_1) = \delta(x_1 + a) - \frac{1}{2} \delta(x_1 + a + b) - \frac{1}{2} \delta(x_1 + a - b).
$$
Here and in what follows we everywhere assume without lost of generality that $F=1$.

%%%%%%%%%%%%%%%%%%%%%%%%%%%%%%%%%%%%%%%%%%%%%%%%%%%%%%%%%%%%%%%%%%%%%%
%%%%%%%%%%%%%%%%%%%%%%%%%%%%%%%%%%%%%%%%%%%%%%%%%%%%%%%%%%%%%%%%%%%%%%

Applying the Mellin transform, we obtain
\beq
\langle \tilde{p} \rangle(s) = \frac{1}{2} a^s + \frac{1}{4} (a + b)^s + \frac{1}{4} (a - b)^s,
\eeq
\beq
\jump{0.15}{\tilde{p}}(s) = a^s - \frac{1}{2} (a + b)^s - \frac{1}{2} (a - b)^s,
\eeq
so that, the zero-order stress intensity factor, corresponding to the unperturbed problem, is given by
\beq
K_\modIII^0 = - \sqrt{\frac{2}{\pi}} \left\{ \frac{1 + \eta}{2} a^{-1/2} + 
\frac{1 - \eta}{4} (a + b)^{-1/2} + \frac{1 - \eta}{4} (a - b)^{-1/2} \right\}.
\eeq

\subsection{Perturbation of crack faces}
\label{sec41}
Let us consider first the perturbation of the crack faces, defined by the two functions $\psi_+$ and $\psi_-$, which have the form
\beq
\psi_\pm(x_1) = \pm \frac{A_\pm}{d_\pm^4} (x_1 + c_\pm + d_\pm)^2 (x_1 + c_\pm - d_\pm)^2,
\eeq
and have supports in the intervals $[-c_\pm - d_\pm, -c_\pm + d_\pm]$. Note that the conditions $c_\pm + d_\pm < a -b$ must be 
satisfied in order to have supports of the functions $\psi_\pm$ separated from the loading.

From eq. \eq{k1a}, we get
\beq
K_\modIII^{1(a)} = \frac{1}{\sqrt{2\pi}} \sum_\pm \frac{1 \pm
\eta}{2}
\int_{-c_\pm - d_\pm}^{-c_\pm + d_\pm} \mu_\pm u_0^{\pm}(x_1)
\frac{\partial}{\partial x_1} \left( (-x_1)^{-3/2} \psi_\pm(x_1)
\right) dx_1,
\eeq
where the term $\mu_\pm u_0^{\pm}(x_1)$ in the integrand can be computed using \eq{int1a}, to obtain
$$
\mu_\pm u_0^{\pm}(x_1) = 
\frac{1 \mp \eta}{2} \left\{ 
I_2(\log\frac{a}{-x_1}) - \frac{1}{2}I_2(\log\frac{a + b}{-x_1}) - \frac{1}{2}I_2(\log\frac{a - b}{-x_1}) 
\right\} \mp
$$
\beq
\left\{
\frac{1 + \eta}{2} I_1(\log\frac{a}{-x_1}) + \frac{1 - \eta}{4}I_1(\log\frac{a + b}{-x_1}) + \frac{1 - \eta}{4}I_1(\log\frac{a - b}{-x_1})
\right\},
\eeq
in which
\beq
I_1(\beta) = \frac{1}{2\pi i} \int_{-i\infty + \delta}^{i\infty + \delta} \frac{\tan(\pi s)}{s} e^{\beta s} ds = 
\frac{1}{\pi} \int_0^\infty \frac{\tanh (\pi t)}{t} \cos(\beta t) dt,
\eeq
\beq
\label{gena1}
I_2(\beta) = \frac{1}{2\pi i} \int_{-i\infty + \delta}^{i\infty + \delta} \frac{\cot(\pi s)}{s} e^{\beta s} ds = 
\frac{\beta}{2\pi} \big(1\pm1\big) \mp 
\frac{e^{\mp\beta/2}}{\pi} \int_0^\infty \frac{\tanh(\pi t)}{t^2 + 1/4} \left(\frac{1}{2}\sin(\beta t) \pm t \cos(\beta t)\right) dt,
\eeq
where the sign depends on whether $\pm\beta>0$.

The integrals $I_1(\beta)$ and $I_2(\beta)$ can be evaluated in closed form (see for instance Gradshteyn and Ryzhik, 2007):
\beq
I_1(\beta) = \frac{1}{\pi} \log \coth \frac{|\beta|}{4},
\eeq
\beq
\label{gena2}
I_2(\beta) = \frac{\beta}{2\pi} + \frac{1}{\pi} \log\left(2\sinh\frac{|\beta|}{2}\right).
\eeq
One can check that
\beq 
I_1(\beta) = -\frac{1}{\pi} \log\frac{|\beta|}{4} +O(\beta^2), \quad \beta \to 0
\eeq
\beq
I_1(\beta) =
\frac{2}{\pi}e^{-|\beta|/2}+O(e^{-|\beta|}), \quad \beta \to \pm\infty
\eeq
\beq
I_2(\beta) = 
\frac{\beta}{2\pi} + \frac{1}{\pi}\log|\beta| + O(\beta^2), \quad \beta \to 0
\eeq
\beq
I_2(\beta) =
\frac{\beta + |\beta|}{2\pi} - \frac{1}{\pi}e^{-|\beta|} + O\left(e^{-2|\beta|}\right), \quad \beta \to \pm\infty.
\eeq
The estimations show that both functions $I_1$ and $I_2$ are singular for small argument values and also 
$I_2$ tends to infinity for large positive values of its argument.

The integral formula for $K_\modIII^{1(a)}$ is
$$
K_\modIII^{1(a)} = 
\frac{1}{\sqrt{2\pi}} \sum_\pm \frac{1 \pm \eta}{2} \cdot 
\int_{-c_\pm - d_\pm}^{-c_\pm + d_\pm} \left\{ 
\frac{1 \mp \eta}{2} \left[ I_2(\log\frac{a}{-x_1}) - \frac{1}{2}I_2(\log\frac{a + b}{-x_1}) - \frac{1}{2}I_2(\log\frac{a - b}{-x_1}) \right] 
\mp \right.
$$
\beq
\label{kiiia}
\left. \left[ \frac{1 + \eta}{2} I_1(\log\frac{a}{-x_1}) + \frac{1 - \eta}{4}I_1(\log\frac{a + b}{-x_1}) + 
\frac{1 - \eta}{4}I_1(\log\frac{a - b}{-x_1}) \right] \right\} 
\frac{\partial}{\partial x_1}\left( (-x_1)^{-3/2} \psi_\pm(x_1) \right) dx_1
\eeq

\begin{figure}[p]
\begin{center}
\includegraphics[width=11.5cm]{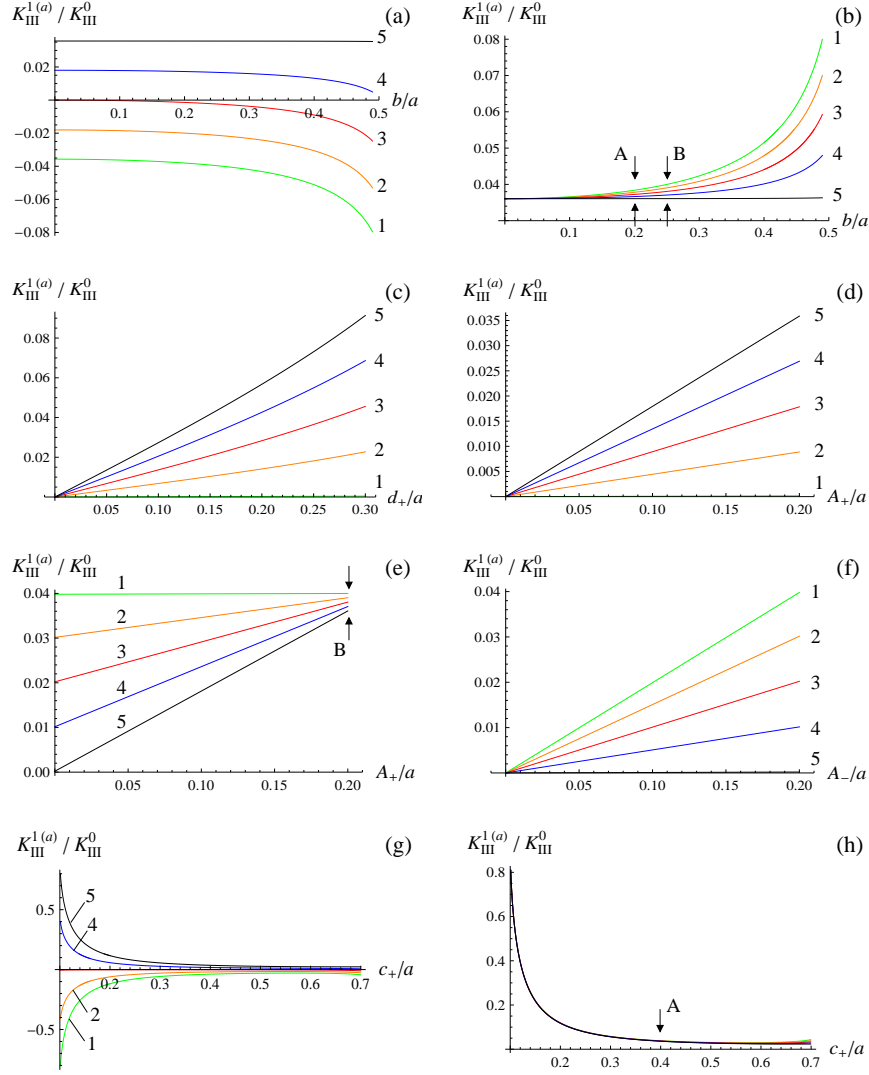}
\caption{\footnotesize Numerical results for perturbation of the crack faces: (1 - green) $\eta = -0.99$,
(2 - orange) $\eta = -0.5$, (3 - red) $\eta = 0$, (4 - blue) $\eta = 0.5$, (5 - black) $\eta = 0.99$.}
\label{fig05}
\end{center}
\end{figure}

Numerical results for different values of parameters are shown in Fig. \ref{fig05}. We note that for the regular perturbation geometry, 
the integrand in  \eq{kiiia} may allow for a singularity not higher than $x_1^{-1/2} \cdot \log x_1$ for admissible values of the 
geometrical parameters $a, b, c_\pm, d_\pm.$
%%%%%%%%%%%%%%%%%%%%%%%%%%%%%%%%%%%%%%%%%%%%%%%%%%%%%%%%%%%%%%%%%%%%%%
%%%%%%%%%%%%%%%%%%%%%%%%%%%%%%%%%%%%%%%%%%%%%%%%%%%%%%%%%%%%%%%%%%%%%%

For parts (a) and (b) of this figure, we use skew-symmetric and symmetric perturbation of the crack faces, respectively. The geometrical 
parameters are shown in Table 1. 
\begin{table}
\label{tab01}
\begin{center}
\caption{\footnotesize Parameters used for plots in Fig. \ref{fig05}.}
\footnotesize
\begin{tabular}{|l|l|l|l|}
\hline
 & Parameters defining the loading & Parameters defining $\psi_+$ & Parameters defining $\psi_-$ \\
\hline
(a) &
$a = 1, \quad 0 < b < 0.5$ &
$A_+ = 0.2, \quad c_+ = 0.4, \quad d_+ = 0.1$ &
$A_- = -0.2, \quad c_- = 0.4, \quad d_- = 0.1$ \\
\hline
(b) &
$a = 1, \quad 0 < b < 0.5$ &
$A_+ = 0.2, \quad c_+ = 0.4, \quad d_+ = 0.1$ &
$A_- = 0.2, \quad c_- = 0.4, \quad d_- = 0.1$ \\
\hline
(c) &
$a = 1, \quad b = 0.2$ &
$A_+ = 0.2, \quad c_+ = 0.5, \quad 0 < d_+ < 0.3$ &
$A_- = 0, \quad c_- = 0.5, \quad d_- = 0.1$ \\
\hline
(d) &
$a = 1, \quad b = 0.25$ &
$0 < A_+ < 0.2, \quad c_+ = 0.4, \quad d_+ = 0.1$ &
$A_- = 0, \quad c_- = 0.4, \quad d_- = 0.1$ \\
\hline
(e) &
$a = 1, \quad b = 0.25$ &
$0 < A_+ < 0.2, \quad c_+ = 0.4, \quad d_+ = 0.1$ &
$A_- = 0.2, \quad c_- = 0.4, \quad d_- = 0.1$ \\
\hline
(f) &
$a = 1, \quad b = 0.25$ &
$A_+ = 0, \quad c_+ = 0.4, \quad d_+ = 0.1$ &
$0 < A_- < 0.2, \quad c_- = 0.4, \quad d_- = 0.1$ \\
\hline
(g) &
$a = 1, \quad b = 0.2$ &
$A_+ = 0.2, \quad 0.1 < c_+ < 0.7, \quad d_+ = 0.1$ &
$A_- = -0.2, \quad c_- = c_+, \quad d_- = 0.1$ \\
\hline
(h) &
$a = 1, \quad b = 0.2$ &
$A_+ = 0.2, \quad 0.1 < c_+ < 0.7, \quad d_+ = 0.1$ &
$A_- = 0.2, \quad c_- = c_+, \quad d_- = 0.1$ \\
\hline
\end{tabular}
\end{center}
\end{table}
The load is produced by point forces applied to the crack surfaces, as shown in Fig. \ref{fig04}. 
A certain asymmetry is introduced in the applied load: one point force is applied on the upper crack face, whereas two smaller point 
forces at shifted positions are applied on the lower crack face; the total load is self-balanced, and the asymmetry is characterised 
by the ratio $b/a$ representing the normalised distance between the two point forces on the lower crack face. Diagrams in 
Fig. \ref{fig05}a,b show the first-order correction in the stress intensity factor for the cases of different asymmetry in the 
applied load and different values of the contrast parameter $\eta = (\mu_--\mu_+)/(\mu_++\mu_-).$ When the perturbation of the crack 
faces is skew-symmetric (see part (a)) the first-order variation in the stress intensity factor becomes zero for the case when the 
materials above and below the crack are equal and when $b/a=0$, as expected. For the case of a bi-material interface, the sign of 
the perturbation depends on the material allocation and the stress intensity factor is shown to be sensitive to the asymmetry in the 
applied load. It is also visible on the diagrams of Figs. \ref{fig05}a and \ref{fig05}b that when the two forces are applied to the 
crack face on the stiffer half-plane (see the curve with the label $5$) the sensitivity to the asymmetry in the applied load becomes 
insignificant. 

In Fig. \ref{fig05}c, we show the first-order correction for the stress intensity factor for the case when the lower face of the crack 
is not perturbed whereas the upper face  has sustained a smooth perturbation with increasing support. It is shown that the increase in 
the size of the perturbation region leads to the increase in the stress intensity factor. It is also noted that the case of a soft upper 
half-plane, corresponding to the perturbed side of the crack, leads to the larger change in the stress intensity factor.

In Figs. \ref{fig05}d and \ref{fig05}e, we show the cases when the lower face of the crack is not perturbed (as in Fig. \ref{fig05}d) 
or perturbed with fixed amplitude (as in Fig. \ref{fig05}e), whereas the upper face is perturbed with the variable amplitude. It is shown 
that the correction for the stress intensity factor (initially zero for case \ref{fig05}d, corresponding to the unperturbed geometry) 
grows linearly as the slope is different for different materials, following the same pattern as in the case \ref{fig05}c. Note that the 
values marked by the arrows in Fig. \ref{fig05}e, corresponding to $A_+/a = 0.2$, are identical to the values marked by the arrows in 
Fig. \ref{fig05}b, corresponding to $b/a = 0.25$, since the two cases correspond to the same geometrical configuration.

Fig. \ref{fig05}f shows the case when the upper face of the crack is not perturbed whereas the lower face is perturbed with the variable 
amplitude, a situation opposite to the case \ref{fig05}d. These computations confirm the conclusion that the variation of the stress 
intensity factor is stronger when the perturbation is applied to the softer material.

In Fig. \ref{fig05}g we show the results obtained for the skew-symmetric perturbation of the crack faces ``moving'' along the crack, 
so that the perturbed boundary becomes close to the region of applied tractions for $c_+/a \to 0.7^-$ and to the crack tip for 
$c_+/a \to 0.1^+$. It is shown that the highest impact on the SIF appears when the perturbation is located near the crack tip, 
whereas a slight increase is observed when the perturbation appears near the acting load. It is also noted that for the homogeneous 
material this contribution is negligibly small. Fig. \ref{fig05}h corresponds to the case when the ``moving'' perturbation of the 
crack faces is symmetric. In this case, the influence of material parameters appears to be small (see also Fig. \ref{fig05}b, where 
the influence of material parameters is enlarged by the refined scale) and the effect on the SIF increases considerably when the 
perturbation is approaching the crack tip, even for homogeneous material.

\subsection{Perturbation of the interface}
\label{sec42}
We consider now the perturbation of the interface, defined by the function $\phi$, which has the form
\beq
\phi(x_1) = \frac{A}{d^4} (x_1 + c + d)^2 (x_1 + c - d)^2,
\eeq
and has support in the interval $[c - d, c + d]$. From eqs. \eq{SIF_1b} and \eq{int1a}, we obtain
$$
K_\modIII^{1(b)} = \frac{F \eta}{2\sqrt{2\pi}} \times
$$
\beq 
\int_{c-d}^{c+d} 
\left\{ 
(1 + \eta) I_3(\log\frac{a}{x_1}) + \frac{1 - \eta}{2} I_3(\log\frac{a + b}{x_1}) + \frac{1 - \eta}{2} I_3(\log\frac{a - b}{x_1}) 
\right\} x_1^{-5/2} \phi(x_1) dx_1, 
\eeq
where
\beq
I_3(\beta) = \frac{1}{\pi} \frac{\sinh\frac{\beta}{2}}{\sinh\beta}.
\eeq

\begin{figure}[ht!]
\begin{center}
\includegraphics[width=11.5cm]{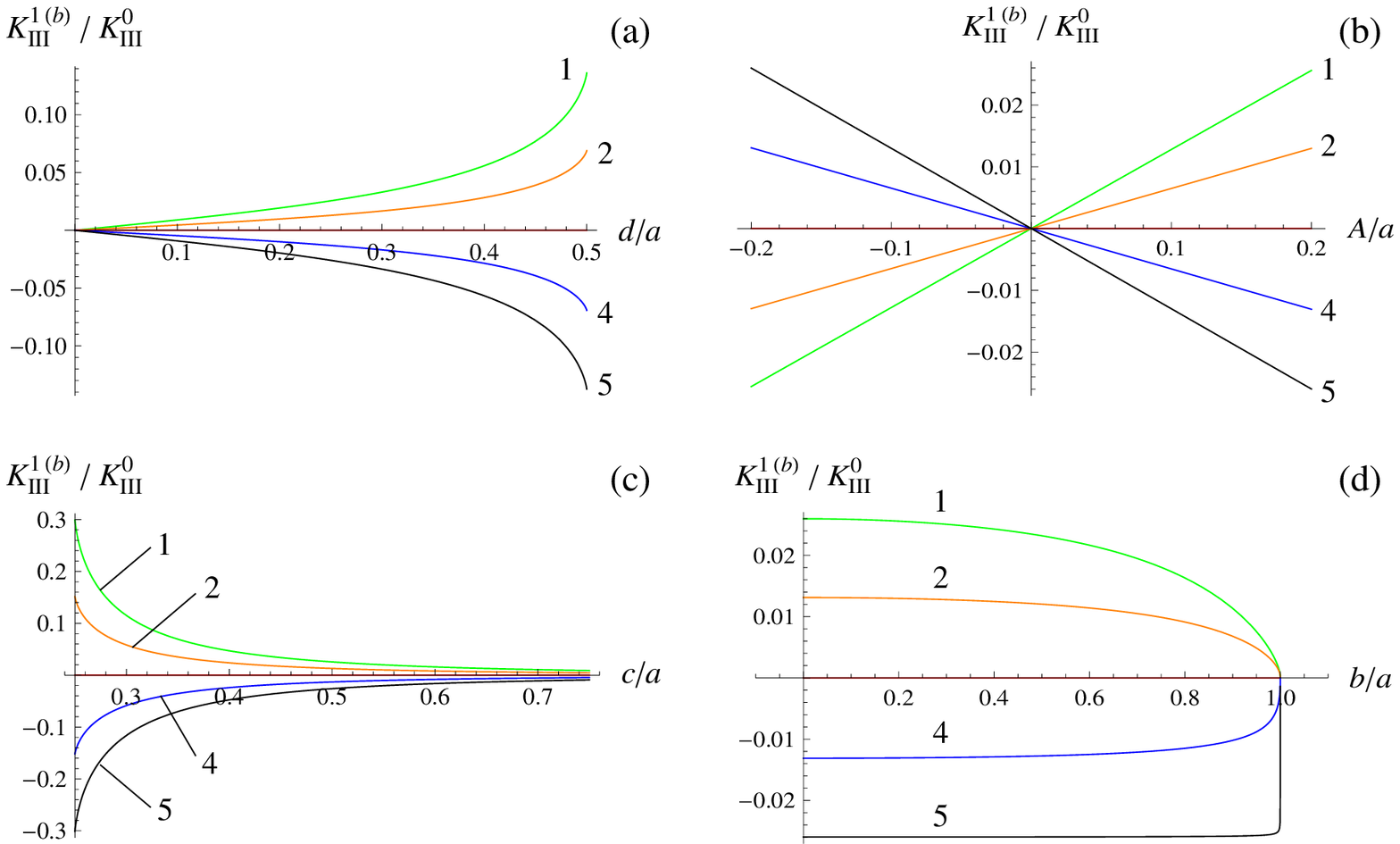}
\caption{\footnotesize Numerical results for perturbation of the interface: 
(1 - green) $\eta = -0.99$, (2 - orange) $\eta = -0.5$, (3 - red) $\eta = 0$, (4 - blue) $\eta = 0.5$, (5 - black) $\eta = 0.99$.}
\label{fig06}
\end{center}
\end{figure}

Numerical computations of the perturbation of the stress intensity factor for different values of parameters are shown in 
Fig. \ref{fig06}. The geometrical parameters characterising the perturbations are given in Table 2. 
%%%%%%%%%%%%%%%%%%%%%%%%%%%%%%%%%%%%%%%%%%%%%%%%%%%%%%%%%%%%%%%%%%%%%%
%%%%%%%%%%%%%%%%%%%%%%%%%%%%%%%%%%%%%%%%%%%%%%%%%%%%%%%%%%%%%%%%%%%%%%
\begin{table}[ht!]
\label{tab02}
\begin{center}
\caption{\footnotesize Parametersused for plots in Fig. \ref{fig06}.}
\footnotesize
\begin{tabular}{|l|l|l|}
\hline
 & Parameters defining the loading & Parameters defining $\phi$ \\
\hline
(a) &
$a = 1, \quad b = 0.2$ &
$A = 0.2, \quad c = 0.5, \quad 0 < d < 0.5$ \\
\hline
(b) &
$a = 1, \quad b = 0.2$ &
$-0.2 < A < 0.2, \quad c = 0.5, \quad d = 0.25$ \\
\hline
(c) &
$a = 1, \quad b = 0.2$ &
$A = 0.2, \quad 0.25 < c < 0.75, \quad d = 0.25$ \\
\hline
(d) &
$a = 1, \quad 0 < b < 1$ &
$A = 0.2, \quad c = 0.5, \quad d = 0.25$ \\
\hline
\end{tabular}
\end{center}
\end{table}

First, we note that the case of $\eta=0$ corresponds to zero perturbation of the stress intensity factor, since this situation corresponds to 
a straight crack in a homogeneous medium. In part (a) of the figure, we show the case when the amplitude of the perturbation is fixed, whereas 
the width of the perturbation region is subject to change. As expected, the increase in the size of the perturbation region leads to the increase 
of the magnitude of the stress intensity factor. If the tractions are applied to the crack faces at a finite distance from the crack tip 
then the asymmetry in the applied load does not give a substantial influence to the magnitude of the perturbation of the stress 
intensity factor. On the part (b) of the figure, we look at the stress intensity factor as a function of the amplitude of the perturbation 
of the interface ahead of the crack. As expected, this dependence is linear, and the asymmetry of the applied load does not give much 
influence on the results of the computations. The diagram (c) of the figure corresponds to the case when the perturbation region of 
a fixed width and amplitude is displaced along the interface ahead of the crack. As expected, when the perturbation region moves towards 
the crack tip the magnitude of the perturbation of the stress intensity factor is increasing. Both cases, (a) and (c) show that the 
perturbation of the stress intensity factor is bounded, as predicted for the case of the regular perturbation of the interface. 
The last diagram (d) in Fig \ref{fig06} illustrates the influence of the asymmetry of the applied load on the perturbation of the 
stress intensity factor. Namely, we consider the extreme situations when one of the forces applied to the lower face of the crack is 
allowed to approach the crack tip. In this case, $K_{III}^{0}$ grows and correspondingly one can see the ratio of 
$K_{III}^{1(b)}/K_{III}^{0}$ decreasing. We also note that the curve with the label $5$ corresponds to the stiff material in the lower 
half-plane, and hence the change in the stress intensity factor is visible only for small distance between the region of loading and 
the crack tip.

\section{Discussions and conclusions}
\label{sec5}
In this paper, we have illustrated the use of the skew-symmetric weight functions in evaluation of the stress intensity factors for 
regularly perturbed interfacial cracks in problems of anti-plane shear. The influence of asymmetry of the applied load as well as 
geometrical perturbations has been shown by the numerical simulations based on the explicit asymptotic formulae. Although the computations 
were presented for the case of a crack in an infinite two-phase plane, the symmetric and skew-symmetric weight functions for a finite 
domain can be constructed by a straightforward superposition of the infinite plane singular solution and an auxiliary solution with the 
finite energy constructed for the finite domain. The asymptotic algorithm for evaluation of the stress intensity factors, presented here, 
equally works for regularly perturbed domains of finite size.

\clearpage
%%%%%%%%%%%%%%%%%%%%%%%%%%%%%%%%%%%%%%%%%%%%%%%%%%%%%%%%%%%%%%%%%%%%%%%
%Appendix
\appendix
\renewcommand{\theequation}{\thesection.\arabic{equation}}

\section{APPENDIX}
\setcounter{equation}{0}

\subsection{Representation in terms of Mellin transform}
\label{app01}

In this section we summarize the solution of the Mode III loading of the interfacial crack, by means of Mellin transform.

The problem is formulated in terms of the Laplace equation
\beq
\Delta u^\pm = \frac{\partial^2 u^\pm}{\partial x_1^2} + \frac{\partial^2 u^\pm}{\partial x_2^2} = 0,
\eeq
subject to the boundary conditions along the crack faces ($x_1 < 0$)
\beq
\left. \mu_\pm \frac{\partial u^\pm}{\partial x_2} \right|_{x_2 = 0^\pm} = p_\pm(x_1),
\eeq
and the transmission conditions along the imperfect interface with presribed discontinuities of displacement and traction ($x_1 > 0$)
\beq
\jump{0.15}{u}(x_1) = u^+(x_1,0^+) - u^-(x_1,0^-) = g_1(x_1), 
\eeq
\beq
\jump{0.15}{\mu \frac{\partial u}{\partial x_2}} = 
\left. \mu_+ \frac{\partial u^+}{\partial x_2} \right|_{x_2 = 0^+} - \left. \mu_- \frac{\partial u^-}{\partial x_2} \right|_{x_2 = 0^-} = 
g_2(x_1).
\eeq

In polar coordinates the Mellin transforms for the displacement vector and the stress tensor with locally bounded elastic energy are defined
as follows
\beq
\tilde{\bu}(s,\theta) = \int_0^\infty \bu(r,\theta) r^{s-1} dr, \quad
\tilde{\bsigma}(s,\theta) = \int_0^\infty \bsigma(r,\theta) r^s dr,
\eeq
and they are represented by analytic functions in the strips $-\vartheta_0 < \Re(s) < \vartheta_\infty$ and $-\gamma_0 < \Re(s) < \gamma_\infty$,
respectively, where $\vartheta_0,\vartheta_\infty \ge 0$ ($\vartheta_0 + \vartheta_\infty > 0$), $\gamma_0,\gamma_\infty > 0$ are constants related
to the behaviour of the solution at the crack tip and at infinity, namely
\beq
\bu(r,\theta) =
\left\{
\barr{ll}
O(r^{\vartheta_0}), & r \to 0, \\
O(r^{-\vartheta_\infty}), & r \to \infty,
\earr
\right.
\quad
\bsigma(r,\theta) =
\left\{
\barr{ll}
O(r^{\gamma_0-1}), & r \to 0, \\
O(r^{-\gamma_\infty-1}), & r \to \infty,
\earr
\right.
\eeq

Correspondingly, the inverse transforms are
\beq
\bu(r,\theta) =
\frac{1}{2\pi i} \int_{\omega_1 - i\infty}^{\omega_1 +i\infty} \tilde{\bu}(s,\theta)
r^{-s} ds, \quad
\bsigma(r,\theta) =
\frac{1}{2\pi i} \int_{\omega_2 - i\infty}^{\omega_2 + i\infty} \tilde{\bsigma}(s,\theta)
r^{-s - 1} ds,
\eeq
where $-\vartheta_0 < \omega_1 < \vartheta_\infty$ and $-\gamma_0 < \omega_2 < \gamma_\infty$.

Applying the Mellin transform to the Laplace equation,
\beq
\Delta u^\pm = \frac{1}{r} \frac{\partial u^\pm}{\partial r} + \frac{\partial^2 u^\pm}{\partial r^2}
+ \frac{1}{r^2} \frac{\partial^2 u^\pm}{\partial \theta^2} = 0,
\eeq
we obtain
\beq
\label{uno}
\frac{\partial^2 \tilde{u}^\pm(s,\theta)}{\partial \theta^2} + s^2 \tilde{u}^\pm(s,\theta) = 0,
\eeq
The general solution of \eq{uno} is given by
\beq
\tilde{u}^\pm(s,\theta) = A^\pm \cos s\theta + B^\pm \sin s\theta.
\eeq
In polar coordinates, the boundary and interface conditions take the form
\beq
-\left. \mu_\pm \frac{1}{r} \frac{\partial u^\pm}{\partial \theta} \right|_{\theta = \pm \pi} = p_\pm(r),
\eeq
\beq
\jump{0.15}{u}(r) = u^+(r,\theta = 0^+) - u^-(r,\theta = 0^-) = g_1(r),
\eeq
\beq
\left. \mu_+ \frac{1}{r} \frac{\partial u^+}{\partial \theta} \right|_{\theta = 0^+} - 
\left. \mu_- \frac{1}{r} \frac{\partial u^-}{\partial \theta} \right|_{\theta = 0^-} = 
g_2(r),
\eeq
and applying the Mellin transform
\beq
-\left. \mu_\pm \frac{\partial \tilde{u}^\pm}{\partial \theta} \right|_{\theta = \pm \pi} = \tilde{p}_\pm(s),
\eeq
\beq
\jump{0.15}{\tilde{u}}(s) = \tilde{u}^+(s,\theta = 0^+) - \tilde{u}^-(s,\theta = 0^-) = \tilde{g}_1(s) = 
\int_{0}^{\infty} g_1(r) r^{s-1} dr,
\eeq
\beq
\left. \mu_+ \frac{\partial \tilde{u}^+}{\partial \theta} \right|_{\theta = 0^+} -
\left. \mu_- \frac{\partial \tilde{u}^-}{\partial \theta} \right|_{\theta = 0^-} = \tilde{g}_2(s) = 
\int_{0}^{\infty} g_2(r) r^s dr,
\eeq
so that
\beq
-\mu_\pm (\mp sA^\pm \sin \pi s + sB^\pm \cos \pi s) = \tilde{p}_\pm(s)
\eeq
\beq
A^+ - A^- = \tilde{g}_1(s),
\eeq
\beq
(\mu_+ B^+ - \mu_- B^-) s = \tilde{g}_2(s).
\eeq

The solution is given by
$$
\tilde{u}^\pm(s,\theta) = \frac{(\tilde{p}_+ - \tilde{p}_- \pm \mu_\mp \tilde{g}_1 s \sin\pi s + \tilde{g}_2 \cos\pi s) \cos s\theta}
{(\mu_+ + \mu_-) s \sin\pi s} +
$$
\beq
\hspace{30mm} \frac{(-\mu_- \tilde{p}_+ - \mu_+ \tilde{p}_- + \mu_+\mu_- \tilde{g}_1 s \sin\pi s \pm \mu_\pm \tilde{g}_2 \cos\pi s) \sin s\theta}
{\mu_\pm (\mu_+ + \mu_-) s \cos\pi s}.
\eeq
Introducing the symmetric and skew-symmetric loading, $\jump{0.15}{\tilde{p}} = \tilde{p}_+ - \tilde{p}_-$ and
$\langle \tilde{p} \rangle = (\tilde{p}_+ + \tilde{p}_-)/2$ respectively, we obtain
\beq
\tilde{u}^\pm(s,\theta) = \left[ \frac{\cos s\theta}{(\mu_+ + \mu_-) s \sin\pi s} +
\frac{(\mu_+ - \mu_-) \sin s\theta}{2\mu_\pm (\mu_+ + \mu_-) s \cos\pi s} \right] \jump{0.15}{\tilde{p}}(s) -
\frac{\sin s\theta}{\mu_\pm s \cos\pi s} \langle \tilde{p} \rangle(s)
\eeq
\beq
+ \left[ \pm \frac{\mu_\mp \cos s\theta}{\mu_+ + \mu_-} + 
\frac{\mu_+\mu_- \sin\pi s \sin s\theta}{\mu_\pm(\mu_+ + \mu_-)\cos\pi s} \right] \tilde{g}_1(s) + 
\left[ \frac{\cos\pi s \cos s\theta}{(\mu_+ + \mu_-) s \sin\pi s} \pm \frac{\sin s\theta}{(\mu_+ + \mu_-) s} \right] \tilde{g}_2(s)
\eeq

In polar coordinates, the non-zero stress components are given by
\beq
\sigma_{3\theta}^\pm(r,\theta) = \mu_\pm \frac{\partial u^\pm}{r\partial \theta}, \quad
\sigma_{3r}^\pm(r,\theta) = \mu_\pm \frac{\partial u^\pm}{\partial r},
\eeq
and after Mellin transform,
\beq
\tilde{\sigma}_{3\theta}^\pm(s,\theta) = \mu_\pm \frac{\partial \tilde{u}^\pm}{\partial \theta} =
\left[ -\frac{\mu_\pm \sin s\theta}{(\mu_+ + \mu_-) \sin\pi s} +
\frac{(\mu_+ - \mu_-) \cos s\theta}{2(\mu_+ + \mu_-) \cos\pi s} \right] \jump{0.15}{\tilde{p}}(s) -
\frac{\cos s\theta}{\cos\pi s} \langle \tilde{p} \rangle(s)
\eeq
\beq
+ \left[ \mp \frac{\mu_+\mu_- s \sin s\theta}{\mu_+ + \mu_-} + 
\frac{\mu_+\mu_- s \sin\pi s \cos s\theta}{(\mu_+ + \mu_-)\cos\pi s} \right] \tilde{g}_1(s) + 
\left[ -\frac{\mu_\pm \cos\pi s \sin s\theta}{(\mu_+ + \mu_-) \sin\pi s} \pm \frac{\mu_\pm \cos s\theta}{(\mu_+ + \mu_-)} \right] \tilde{g}_2(s)
\eeq
\beq
\tilde{\sigma}_{3r}^\pm(s,\theta) = -s \mu_\pm \tilde{u}^\pm =
- \left[ \frac{\mu_\pm \cos(s\theta)}{(\mu_+ + \mu_-) \sin(\pi s)} +
\frac{(\mu_+ - \mu_-) \sin(s\theta)}{2(\mu_+ + \mu_-) \cos(\pi s)} \right] \jump{0.15}{\tilde{p}}(s) +
\frac{\sin(s\theta)}{\cos(\pi s)} \langle \tilde{p} \rangle(s)
\eeq
\beq
- \left[ \pm \frac{\mu_+\mu_- s \cos s\theta}{\mu_+ + \mu_-} + 
\frac{\mu_+\mu_- s \sin\pi s \sin s\theta}{(\mu_+ + \mu_-)\cos\pi s} \right] \tilde{g}_1(s) - 
\left[ \frac{\mu_\pm \cos\pi s \cos s\theta}{(\mu_+ + \mu_-) \sin\pi s} \pm \frac{\mu_\pm \sin s\theta}{(\mu_+ + \mu_-)} \right] \tilde{g}_2(s)
\eeq

By means of the same procedure as in Piccolroaz et al. (2009), we obtain the asymptotics of stress and displacement near the crack tip as follows
\beq
\bsigma(r,\theta) = \sum_{k=1}^N \Res[\tilde{\bsigma}(s,\theta),s=-k/2] r^{k/2-1} + O(r^{\frac{N-1}{2}}),
\eeq
\beq
\bu(r,\theta) = \sum_{k=0}^N \Res[\tilde{\bu}(s,\theta),s=-k/2] r^{k/2} + O(r^{\frac{N+1}{2}}).
\eeq
Since
\beq
\tilde{\sigma}_{3\theta}^\pm(s,0) = 
-\frac{1}{\cos\pi s} \left\{ \langle \tilde{p} \rangle(s) + \frac{\eta}{2} \jump{0.15}{\tilde{p}}(s) \right\} + 
\frac{\mu_+\mu_- s \sin\pi s}{(\mu_+ + \mu_-)\cos\pi s} \tilde{g}_1(s) \pm 
\frac{\mu_\pm}{(\mu_+ + \mu_-)} \tilde{g}_2(s),
\eeq
\beq
\jump{0.15}{\tilde{u}}(s) = -\left(\frac{1}{\mu_+} + \frac{1}{\mu_-}\right) \frac{\sin\pi s}{s \cos\pi s} 
\left\{ \langle \tilde{p} \rangle(s) + \frac{\eta}{2} \jump{0.15}{\tilde{p}}(s) \right\} +
\frac{\tilde{g}_1(s)}{\cos\pi s},
\eeq
the two-terms asymptotics of tractions ahead of the crack tip and crack opening read
\beq
\sigma_{3\theta}(r,0) = \frac{K_\modIII}{\sqrt{2\pi}} r^{-1/2} + \frac{A_\modIII}{\sqrt{2\pi}} r^{1/2} + 
\frac{B_\modIII}{\sqrt{2\pi}} r^{3/2} + O(r^{5/2}),
\eeq
\beq
\jump{0.15}{u}(r) = \frac{2 K_\modIII}{\sqrt{2\pi}} \left( \frac{1}{\mu_+} + \frac{1}{\mu_-} \right) r^{1/2} -
\frac{2 A_\modIII}{3\sqrt{2\pi}} \left( \frac{1}{\mu_+} + \frac{1}{\mu_-} \right) r^{3/2} + 
\frac{2 B_\modIII}{5\sqrt{2\pi}} \left( \frac{1}{\mu_+} + \frac{1}{\mu_-} \right) r^{5/2} + O(r^{7/2}),
\eeq
respectively, where
\beq
K_\modIII = -\sqrt{\frac{2}{\pi}} \int_0^\infty \left\{ \langle p \rangle(r) + \frac{\eta}{2} \jump{0.15}{p}(r) \right\} r^{-1/2} dr + 
\frac{1}{\sqrt{2\pi}} \frac{\mu_+\mu_-}{\mu_+ + \mu_-} \int_{0}^{\infty} g_1(r) r^{-3/2} dr,
\eeq
\beq
A_\modIII = \sqrt{\frac{2}{\pi}} \int_0^\infty \left\{ \langle p \rangle(r) + \frac{\eta}{2} \jump{0.15}{p}(r) \right\} r^{-3/2} dr + 
\frac{3}{\sqrt{2\pi}} \frac{\mu_+\mu_-}{\mu_+ + \mu_-} \int_{0}^{\infty} g_1(r) r^{-5/2} dr,
\eeq
\beq
B_\modIII = -\sqrt{\frac{2}{\pi}} \int_0^\infty \left\{ \langle p \rangle(r) + \frac{\eta}{2} \jump{0.15}{p}(r) \right\} r^{-5/2} dr +
\frac{5}{\sqrt{2\pi}} \frac{\mu_+\mu_-}{\mu_+ + \mu_-} \int_{0}^{\infty} g_1(r) r^{-7/2} dr.
\eeq

\end{document}

%% file: definitions.tex
\newcommand{\beq}{\begin{equation}}
\newcommand{\eeq}{\end{equation}}

\newcommand{\beqar}{\begin{eqnarray}}
\newcommand{\eeqar}{\end{eqnarray}}
\newcommand{\bit}{\begin{itemize}}
\newcommand{\eit}{\end{itemize}}
\newcommand{\benum}{\begin{enumerate}}
\newcommand{\eenum}{\end{enumerate}}
\newcommand{\barr}{\begin{array}}
\newcommand{\earr}{\end{array}}
\def\ds{\displaystyle}
\newcommand\eq[1]{(\ref{#1})}
\newcommand{\jump}[2]{[\mbox{\hspace{-#1em}}[#2]\mbox{\hspace{-#1em}}]}
\newcommand{\bjump}[2]{\left[\mbox{\hspace{-#1em}}\left[#2\right]\mbox{\hspace{-#1em}}\right]}
\newcommand{\modIII}{\text{III}}

\def\XXint#1#2#3{{\setbox0=\hbox{$#1{#2#3}{\int}$}
   \vcenter{\hbox{$#2#3$}}\kern-.5\wd0}}

\def\b0{\mbox{\boldmath $0$}}

\def\bu{\mbox{\boldmath $u$}}

\def\bx{\mbox{\boldmath $x$}}

\newcommand{\bsigma}{\mbox{\boldmath $\sigma$}}

\def\f0{\ensuremath{\mathbb{O}}}

\def\Re{\mathop{\mathrm{Re}}}

\newcommand{\Res}{\mathop{\mathrm{Res}}}

\def\IJSS{{\it Int.\ J.\ Solids Struct.}\ }

\def\JAM{{\it ASME J.\ Appl.\ Mech.}\ }

\def\JMPS{{\it J.\ Mech.\ Phys.\ Solids}\ }